\documentclass[]{spie}  

\usepackage{amsmath,amsfonts,amssymb}
\usepackage{graphicx}
\usepackage[colorlinks=true, allcolors=blue]{hyperref}

\title{Addressing environmental and atmospheric challenges for capturing high-precision thermal infrared data in the field of astro-ecology}

\author[a]{Claire Burke}
\author[a]{Maisie F. Rashman}
\author[b]{Owen McAree}
\author[c]{Leonard Hambrecht}
\author[a]{Steve N. Longmore}
\author[c]{Alex K. Piel}
\author[c,d]{Serge A. Wich}
\affil[a]{Astrophysics Research Institute, LJMU, IC2 Liverpool Science Park, Liverpool, L3 5RF, UK.}
\affil[b]{Faculty of Science, Liverpool John Moores University, James Parsons Building, Byrom Street, L3 3AF, Liverpool, UK.}
\affil[c]{School of Natural Sciences and Psychology, Liverpool John Moores University, James Parsons Building, Byrom Street, L3 3AF, Liverpool, UK.}
\affil[d]{Institute for Biodiversity and Ecosystem Dynamics, University of Amsterdam, Sciencepark 904, Amsterdam 1098, Netherlands.}

\authorinfo{Further author information: (Send correspondence to C.B)\\C.B: E-mail: C.Burke@ljmu.ac.uk}

\begin{document} 
\maketitle

\begin{abstract}
Using thermal infrared detectors mounted on drones, and applying techniques from astrophysics, we hope to support the field of conservation ecology by creating an automated pipeline for the detection and identification of certain endangered species and poachers from thermal infrared data. We test part of our system by attempting to detect simulated poachers in the field. Whilst we find that we can detect humans hiding in the field in some types of terrain, we also find several environmental factors that prevent accurate detection, such as ambient heat from the ground, absorption of infrared emission by the atmosphere, obscuring vegetation and spurious sources from the terrain. We discuss the effect of these issues, and potential solutions which will be required for our future vision for a fully automated drone-based global conservation monitoring system.

\end{abstract}


\section{INTRODUCTION}
\label{sec:intro}  

The rate of decline of species of life on Earth is currently at the highest since the event that wiped out the dinosaurs \cite{Ceb15}. Additionally, in the past 30 years the numbers of wild animals within many individual species has halved \cite{wwf16}. Large scale conservation efforts are required, but effective conservation strategy cannot be formulated without a good understanding of current species distributions and density \cite{Schwarz08,Fre12,McM14,Buc01}. Wildlife surveys are traditionally carried out on foot, by car or by manned aircraft; methods which are costly, time consuming and prone to large uncertainties \cite{Franklin10,Buc01,Schwarz08}. Poaching is also a major threat to endangered animals, however detecting poachers in real time is challenging.

Using drones (a.k.a. unmanned areal vehicles) provides an advantage over ground surveys as drones can cover large areas quickly, and drones are cheaper and safer to deploy than manned aircraft \cite{vermeulen13}. Drones have been used for conservation surveys and some success has been reported in detecting animals with optical cameras \cite{Wich18, hodgeson18}. Optical cameras have the disadvantage of being reliant on sunlight to be useful, hence blind at night when animal activity is high and most poaching occurs. In optical footage all objects appear the same brightness, making automated detection of animals challenging, especially considering that many animals are naturally camouflaged in their environments when viewed at optical wavelengths. The addition of thermal infrared (TIR) cameras allows great potential for automated detection and classification systems for animals in the wild, thus increased accuracy of detections and classifications. In TIR footage warm objects such as animals and humans appear as bright sources. We aim to apply techniques commonly used in astrophysics to detect and classify these sources automatically. 

Using an interdisciplinary approach, the astro-ecology project aim is to help increase the efficiency of species monitoring for conservation and the efficacy of poaching prevention. The end goal of the astro-ecology project is to produce a fully automated detection and classification pipeline for animals and poachers which can be applied in the wild anywhere in the world. The outline of the system we are developing looks like this;
\begin{itemize}
\item Autonomous drone flight over region of interest
\item Detect animals as drone is flying
\item Identify animals using machine learning approach, based on species' unique thermal profile
\item Notify system user (e.g. game warden, conservationist) of locations and identities of animals or poachers as they are detected.
\end{itemize}
See \cite{Longmore17} for more details.

Following the success of our pilot study \cite{Longmore17} we test the feasibility of using our proposed setup to detect poachers in the field with this study. We simulate poachers in the field with students hiding in the bush at a forest reserve in Tanzania. We find that by setting a simple threshold in temperature it is easy to detect the `poachers' in many cases. However, in some cases they are not detected or there are many sources of false detections. We discuss these difficult cases and what would need to be done to resolve them.

\section{DATA}
Data were taken during March 2017 at Tongwe West Joint District-Village Forest Reserve, Tanzania. To simulate poachers a group of students stood or sat in the bush throughout the area surveyed. The area is a mixed-tree species (miombo) woodland of varying density of vegetation, ranging from wide open to partially wooded but without any full or dense canopy cover. The students' locations were chosen to be under differing levels of vegetation coverage, from completely open to under the trees. For each flight, the drone was flown overhead in a grid pattern at 70 meters and 100 meters above ground level. For all flights the grid contained 3 transects, each 70 meters long, and each transect was separated by 10 meters. This small separation between transects allows for significant overlap in the field of view of the camera, meaning edge effects in the camera detector can be accounted for.

Two different sites in the reserve were surveyed in this manner, each with variations on local terrain and vegetation. The sites were close to latitude: $-5.5$, longitude: $30.56$ (full co-ordinates withheld for conservation purposes). Flights were carried out over 4 dates at varying times of day, the details of the flights are shown in Table~\ref{tab:Flight details}. See \cite{hambract18} for full details of the observations performed \footnote{A video of the data being taken can be found here https://youtu.be/p8rmXzOmRfM.}.

Data were taken with a FLIR Tau 640 thermal infrared camera with 13mm lens mounted on a DJI F550 hexacopter drone with the camera aimed straight down (nadir position). The detector has 640 $\times$ 512 pixels and an angular field of view of 45$^{\circ} \times 37^{\circ}$. The angular field of view is related to physical field of view, $FOV$, (ie. instantaneous footprint on the ground) by,
\begin{equation}
\label{equ:fov}
FOV=2h.tan(\theta/2),
\end{equation}
where $h$ is the height of the drone and $\theta$ is the angular field of view.

Figure~\ref{fig:good data} shows some example still frames from the footage taken at different sites and 70 meter and 100 meter heights. As can be seen in the figures, there is some vignetting in the images, where the edges of the field of view report a lower temperature than at the centre. This is due to the optics of the camera. To mitigate the effect of this we ignore the outer 50 pixels in our analysis. Since the transects of the flights were closely spaced we still have significant overlap in the field of view and thus sample the full area surveyed. For reliable detection and analysis of data a detailed understanding of the detector will be required. This is discussed in a separate paper \cite{rashman18}.

\section{METHODS}

We count the number of `poachers' in each flight using a simple thresholding algorithm.
Two different thresholds were applied to filter each image for potential humans.  First, a threshold was applied to the temperature values to select only the pixels which were in the 99th percentile, corresponding to the hottest pixels in the image (discussed below). Secondly, a standard deviation filter was applied to an $n \times n$ kernel surrounding each pixel to provide a measure of contrast in the neighbourhood. A threshold is then applied to this contrast data and only pixels with a high neighbourhood contrast are selected. The kernel size, $n$, is determined by the flying height of the drone so as to correspond to the approximate pixel size of a human candidate. The expected size of humans when viewed from above, and thus the size range for detecting candidate humans was set to be between 20--50 cm. This range allows for viewing from different angles and partial obscuration by objects nearby (e.g. vegetation). Size thresholds in pixels were calculated by simple geometric calculations based on camera resolution, angular field of view and height of drone, see Equation~\ref{equ:fov}.
The binary images produced by these two thresholds were combined into a single image by applying an element-wise \emph{AND} operation. This final binary image is then searched for connected areas of pixels which satisfy the size constraints, and the centroids of these areas extracted as the pixel location of candidate humans.

Warm-blooded (homeothermic) animals can regulate their body temperature. In order to cool themselves, the skin (or surface) temperature of an animal must be warmer than its surroundings. The major advantage of observing using TIR is that homeothermic animals will usually appear warmer than their surroundings. To regulate temperature in this way an animal's skin temperature must change throughout the day to best suit its surroundings, and the animal's need to keep warm or cool down. The consequence of this is that there is no specific characteristic temperature which will always indicate an animal or human - rather a temperature difference between the animal and its surroundings. As such, setting an absolute temperature threshold above which to detect animals is not a reliable method to guarantee detections. The surroundings of the animal can also change temperature, and during the heat of the day may also be above a given absolute detection threshold (discussed further below). An animal's skin temperature can drop below a given threshold as its surroundings cool and its need to dissipate heat drops. (Conversely in very cold environments, where animals need to be well insulated to retain body heat, they can appear the same temperature or even colder than their surroundings \cite{polar_bears}.) For this study we set a relative detection threshold - only selecting objects in the 99th percentile of temperatures in the footage as potential target detections. 
This percentile was determined by preliminary examination of the data. We made a histogram of the temperature of all pixels in the data and found that 95\% of the pixels containing target humans had temperature in the 99th percentile of this histogram. This threshold allows the elimination of most sources of noise and spurious sources whilst minimizing loss of desirable targets.

Each candidate human is tracked from one frame to the next by applying a Kalman Filter (KF) on the pixel locations in order to estimate the inter-frame motion. The KF state predictions are used to determine if a candidate in a subsequent image is likely to correspond to a previously detected candidate by applying a nearest-neighbour cost function. If such a match is found then the KF is updated with this new measurement information. If no such match is found, tracking of that candidate ceases. New candidates which do not correspond to KF predictions from the previous frame are used as the basis of new KF. The total candidate count in a data set corresponds to the total number of KF instances created (values in brackets in Table~\ref{tab:Flight details}). 

We examined all the candidate humans by eye to confirm the numbers of true detections of humans and noted possible reasons for false detections and missed detections of humans. The locations of the humans was recorded by GPS on the ground at the time of data being gathered. These locations were used to confirm true positive and true negative detections.

\section{RESULTS}

\begin{figure} [ht]
\begin{center}
\begin{tabular}{c} 
\includegraphics[height=5cm]{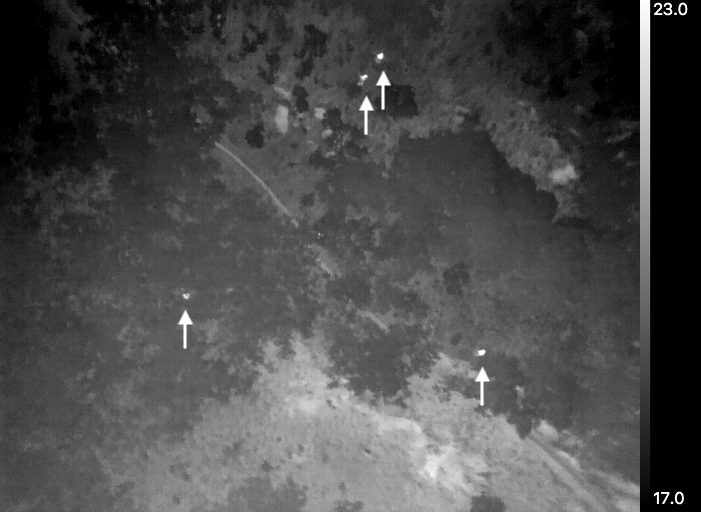}
\includegraphics[height=5cm]{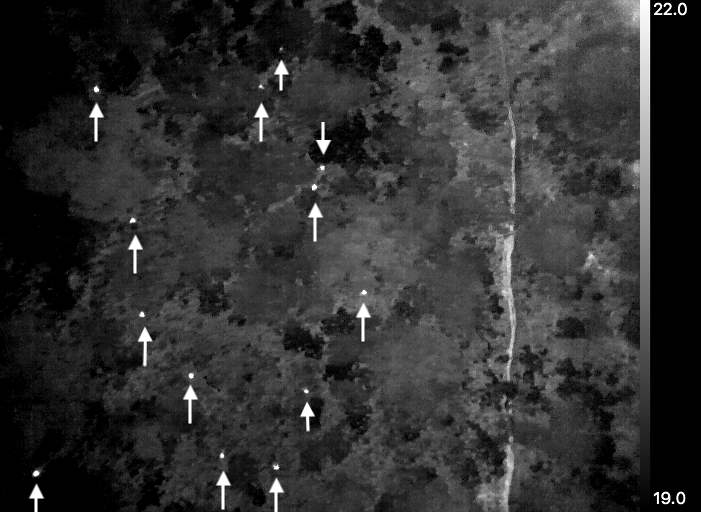}
\end{tabular}
\end{center}
\caption[example] 
{\label{fig:good data} Examples of data where humans are clearly identifiable by eye and successfully detected by the algorithm. Arrows indicate humans. Left: from flight 6 at 70 meters - 4 humans present, FOV=$58\times47$ meters . Right: from flight 4 at 100 meters - 13 humans indicated with arrows, FOV=$82\times 67$ meters. Temperature scale in $^{\circ}$C is indicated on the right of each image. }
\end{figure}

\begin{figure} [ht]
\begin{center}
\begin{tabular}{c} 
\includegraphics[height=5cm]{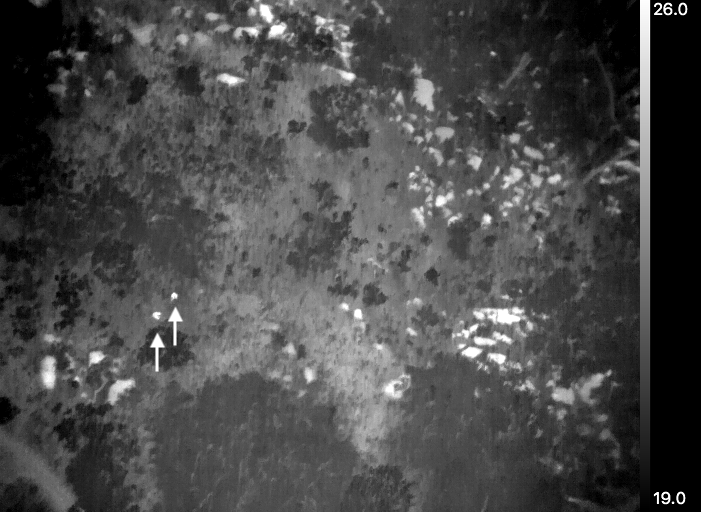}
\includegraphics[height=5cm]{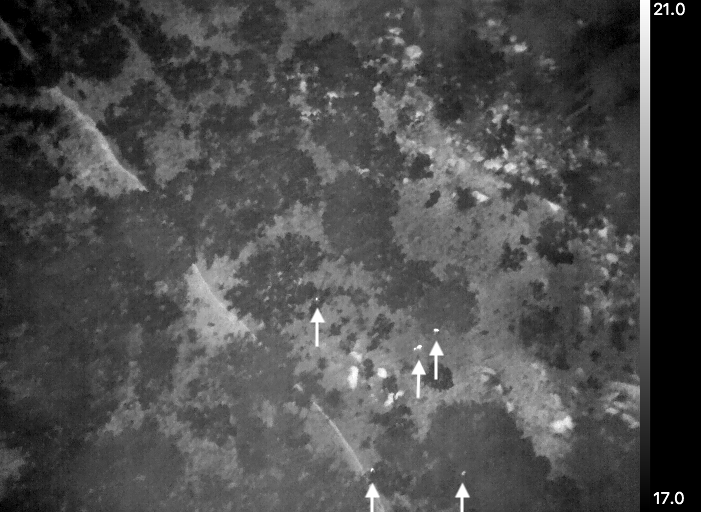}
\end{tabular}
\end{center}
\caption[example] 
{\label{fig:rocks} Examples of data where rocks are a source of error. Arrows indicate humans. Left: Flight 5 at 70 meters - 2 humans, FOV=$58\times47$ meters. Right: Flight 6 at 100 meters - 5 humans, FOV=$82\times 67$ meters.}
\end{figure}

\begin{figure} [ht]
\begin{center}
\begin{tabular}{c} 
\includegraphics[height=5cm]{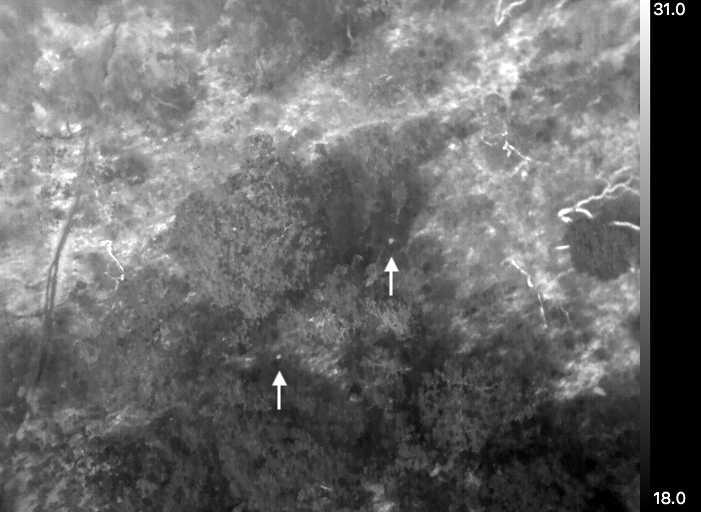}
\includegraphics[height=5cm]{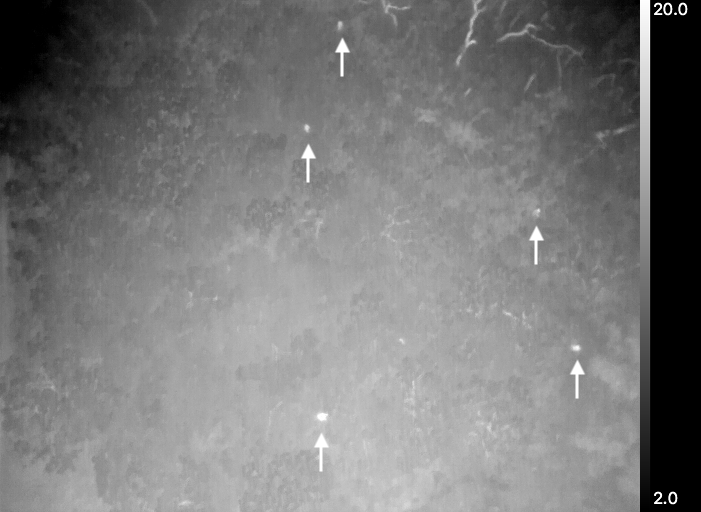}
\end{tabular}
\end{center}
\caption[example] 
{\label{fig:other enviro} Examples of data with other environmental issues. Arrows indicate humans. Left: flight 1 at 70 meters, vegetation obscuring humans, tree branches very reflective providing false detections, the ground is a comparable temperature to humans - 2 targets. Right: flight 2 at 70 meters, data appears hazy - 5 humans. FOV=$58\times47$ meters for both.}
\end{figure}

Table~\ref{tab:Flight details} shows the total numbers of humans correctly and incorrectly counted, and sources of false positive detections or false negative detections (missed humans). The algorithm was able to detect the humans, they also appear fairly clear to the eye in much of the data. However there were a number of targets that were not detected, either by eye or by the algorithm. There were also sources of false positive detections for the algorithm, which can be differentiated by eye.
Figure~\ref{fig:good data} shows data from flights where humans were more reliably detected. Figures~\ref{fig:rocks} and \ref{fig:other enviro} show data from flights where true positive detections were fewer and false positive detections were more numerous.

The causes of false negative (missed) detections were identified as vegetation cover, blending of sources and background (low thermal contrast), confusion with other sources such as hot rocks where humans were close to or on top of the rocks, and where the human temperature was lower than the 99th percentile of the surroundings (e.g. left hand side of Figure~\ref{fig:other enviro}, the two humans indicated were not detected by the algorithm). Causes of false positive detections were hot patches of ground, hot rocks, hot or reflective tree branches. 

\begin{table}[ht]
\caption{Details of flights performed, number of humans in the area surveyed, number of humans detected for each height, number of true positive detections in {\bf bold}, percentage of humans detected in square brackets, total detections in round brackets; and the main sources of false positives or false negatives as a result of the environment excluding vegetation cover which was an issue for all flights.} 
\label{tab:Flight details}
\begin{center}       
\begin{tabular}{|l|l|c|c|c|c|c|c|} 
\hline
\rule[-1ex]{0pt}{3.5ex}  Flight& Site  & Date & Take-off and & \#humans  & \multicolumn{2}{c}{\# detected} & Sources of\\
\rule[-1ex]{0pt}{3.5ex}  number&       &      & landing time &  &  \multicolumn{2}{c}{{\bf True positive} [percentage]} & error \\
\rule[-1ex]{0pt}{3.5ex}  &       &      & & &  \multicolumn{2}{c}{(Total number detected) }   & (excluding \\
\rule[-1ex]{0pt}{3.5ex}  &       &      & & & 70m & 100m  & vegetation cover)\\

\hline
\rule[-1ex]{0pt}{3.5ex}  1 & 1& 20/03/2017 & 16:06, 16:12& 23 & {\bf 3} [13\%] (187) & {\bf 3} [13\%] (117) &  High ground  \\ 
\rule[-1ex]{0pt}{3.5ex}   &&  & & & & &  temperature, \\
\rule[-1ex]{0pt}{3.5ex}   &&  & & & & &  tree branches \\
\rule[-1ex]{0pt}{3.5ex}  2 & 1& 20/03/2017 & 18:46, 18:51& 15 & {\bf 9} [60\%] (44) & {\bf 8} [53\%] (57) & Haze, \\ 
\rule[-1ex]{0pt}{3.5ex}   &&  & & & & &  tree branches\\
\rule[-1ex]{0pt}{3.5ex}  3 & 1& 21/03/2017 & 19:02, 19:08& 15 & {\bf 10} [67\%] (11) & {\bf 11} [73\%] (19) & Hot rocks\\ 
\rule[-1ex]{0pt}{3.5ex}  4 & 1& 21/03/2017 & 07:32, 07:38& 22 & {\bf 17} [77\%] (31) & {\bf 18} [81\%] (26) & Veg only \\ 
\rule[-1ex]{0pt}{3.5ex}  5 & 2& 22/03/2017 & 19:12, 19:19& 25 & {\bf 7} [28\%] (289) & {\bf 7} [28\%] (431) & Hot rocks\\ 
\rule[-1ex]{0pt}{3.5ex}  6 & 2& 23/03/2017 & 07:23, 07:29& 11 & {\bf 11} [100\%] (15) & {\bf 11} [100\%] (11) & Veg only\\ 
\rule[-1ex]{0pt}{3.5ex}  7 & 2& 23/03/2017 & 19:12, 19:18& 7  & {\bf 3} [42\%] (225) & {\bf 4} [57\%] (345) & Branches, \\ 
\rule[-1ex]{0pt}{3.5ex}   &&  & & & & &  hot rocks\\
\hline 
\end{tabular}
\end{center}
\end{table}

\section{DISCUSSION}

Humans were detected by the automated algorithm in all flights, however the percentage of humans recovered was very variable between flights at different sites and times of day. Given the small number of flights for each different time of day it would be difficult to draw robust statistical conclusions from this dataset on the specific dependence of detection on different environmental factors. It is possible with this data to make some qualitative statements about environmental influences on the numbers of both true and false detections which are consistent with physical arguments.

The results in Table~\ref{tab:Flight details} suggest that morning was the best time of day for maximum true positives - with 77\%, 81\% and 100\% of humans detected. In these cases only the vegetation was identified as a problem. 
The worst case for true positives was the mid-afternoon flight (flight 1), when only 13\% of humans were recovered and the number of false positives was very high. 
On the evening flights between 28 and 73\% of humans were detected. There is quite a large variation in success rate for the evening flights. Early in the morning the ground and other objects of terrain and vegetation will be cooler than in the heat of the day when the sun is close to overhead. This means that there will be a larger difference in temperature between humans and their surroundings in the morning making them easier to detect than later in the day. 
Since no detailed record was made of weather conditions, air temperature, cloud cover etc, it is not possible to say if the variation in results at similar times of day was mainly due to changes in solar heating of the ground and surroundings or the different positioning of the `poachers' within the vegetation in different flights.

The highest number of false positives occurred when rocks and tree branches were clear in the data providing spurious sources.
The heating of the ground and other surfaces throughout the day is the cause of this, and the heat absorbed by these surfaces during the day may be retained for several hours after peak solar influx (as was clear in the data). For minimizing false positives it is clear that early morning is the better time to perform observations, however a detailed analysis of the changes in surface temperatures throughout the day would be required to determine the optimum time for observing.

Changing drone flying height did not make any meaningful difference in the number of humans detected for this study. 
For these observations no detailed pre-planning was given to necessary apparent human size within the data for accurate detections. Whilst the change in apparent human size between 70m and 100m drone height above ground level is not significant enough to affect detections, it may be that at lower heights, and hence larger apparent human size, more humans are successfully detected. The optimum height for maximum reliable detections will be dependent on the resolution and sensitivity of the camera. In a real life application of this technology there will also be a trade-off between FOV on the ground and apparent human size, so surveys over larger areas will  need to be optimized to cover the largest possible area quickly whilst still retaining reliable detections.

In Flight 2, which was performed in the evening, the images appear obscured by some kind of haze. Observers on the ground did not report seeing any haze or mist at the time of flying/filming. Since no observations of the local temperature or humidity were recorded, and there is no weather station nearby to check the recorded conditions, it is not possible to say if this effect might have been due to local atmospheric conditions. The origin of this haze is suspected to be condensation on the lens of the camera, assuming the camera was cooler than the air at time of flight and that the atmospheric air contained sufficient moisture. It is not clear what the effect of this haziness was on humans counted.

Whilst only the number of true positives detected is reported in the table in bold, the total number counts (in brackets) has not been adjusted for any humans or other objects that were counted twice. 
Tracking points across multiple frames without any a-priori knowledge of the inter-frame motion requires this motion to be estimated, in this case we have employed a Kalman Filter to do this. Such an approach is well suited to estimating smooth, slowly-changing, motion of either the camera or target. However, this type of motion estimation is poorly suited to jerky movements or situations where both the camera and target are moving, such as was recorded in the data examined here. To alleviate these tracking problems it is desirable to utilize a gimbal-stabilized camera to decouple its motion from that of the drone and include a measure of the drones position in each frame to decouple its motion from that of the target - human or otherwise. This will help to alleviate double counting of potential objects and allow more accurate detections to be obtained.


The major environmental issues we have found with this field test can be summarized as;
\begin{itemize}
\item Ground temperature - during and for several hours after solar noon the temperature of the ground can be close to that of humans. 
\item Spurious heat sources - rocks and tree branches that have been heated by the sun, some tree branches are also very reflective to TIR wavelength light.
\item Vegetation cover over humans.
\item Small apparent size of humans, similarity of size of humans with other objects when viewed from above.
\item Haze, possibly as a result of condensation.
\item Jerky motion of the camera making objects difficult to track, leading to double counting.
\end{itemize}
Potential steps to resolving these issues are;
\begin{itemize}
\item Understanding and modeling the temperature of the ground, potentially allowing background subtraction.
\item Optimization of flight plan with respect to ground temperature expected for a given time of day and expected temperatures of humans or other objects.
\item Modeling terrain and vegetation to preempt and discard spurious sources (will be discussed in a future paper - Burke et al 2019)
\item Optimizing drone flight plan with respect camera resolution for detection based on expected apparent sizes of objects as a function of drone height.
\item Understanding how terrain, ground temperature and vegetation affects what temperature humans and other objects will appear - especially at the edges of sources where the emission recorded per pixel will be blended with that from the background (camera and environmental information required).
\item Understanding of absorption properties of atmosphere (including haze/fog) at thermal infrared wavelengths - the details of air temperature etc were not recorded for the data used here.
\item Controlling the housing and component temperatures of the thermal camera, and the difference with that in the local environment.
\item Camera stabilization by use of a gimbal or similar device, and GPS (or similar) tagging of individual frames in footage.
\end{itemize}
These solutions will be discussed in detail in an upcoming paper \cite{burke18}.

\section{CONCLUSIONS AND FUTURE WORK}

Detecting humans for monitoring and prevention of poaching is not straightforward. Gains in accuracy and area surveyed can potentially be improved with the use of areal footage compared to ground surveying on foot or by vehicle. However even though the addition of thermal equipped drones seems like it should make things easier, it comes with its own challenges. In this paper we have focused on the challenges of observing environment, however there are also challenges associated with the thermal camera itself, which will be discussed in an accompanying SPIE paper \cite{rashman18}. We will present some solutions to the problems presented here in an upcoming publication \cite{burke18}.

Astro-ecology aims to facilitate a step change in the efficiency and efficacy of wildlife monitoring and poaching prevention for conservation. In this paper we have shown that humans posing as poachers can be detected with thermal infrared cameras mounted on drones. However reliably detecting humans in different terrains and vegetation is subject to several environmental challenges - this will also hold true for detecting animals.

To resolve these issues we will need to gain a more thorough understanding of how different observing environments will affect the data gathered, and develop methods and models to account for these factors. This will be vital for future components of our system in development, for example the effectiveness of a machine learning component which will be able to identify targets of interest and throw out spurious sources which are a result of the environment. The observations presented in this paper represent a step towards the astro-ecology goals and will inform future observing strategies and algorithm development going forwards.

\appendix
\acknowledgments 
 
The authors thank the Tanzanian Wildlife Research Institute (TAWIRI) and Commission for Science and Technology (COSTECH) as well as the UCSD/Salk Institute Center for Academic Research and Training in Anthropogeny (CARTA) for support to GMERC.
We thank Andy Goodwin and Ian Thomson for preparing the drone and camera anti-vibration plate. 
SW thanks WWF Netherlands for financial support. 
We gratefully acknowledge funding from the Research Council UK, Science and Technology Facilities Council grant ST/R002673/1.

Special thanks goes to the students of the 2016/2017 LJMU M.Sc. course ``Wildlife Conservation \& UAV Technology'', and the GMERC (formerly Ugalla Primate Project) staff for their support and patience. The authors would like to thank: Claire Rigby, Finnoula Taylor, Megan Melia, Jospeh Goode, Derek Dwane, Naomi Jones, Rory Andrews, Anna Starkey, Joseph Phillips, Naomi Davies, Molly Frost, Evie Hyland, Olivia Evans, Jade Musto, Andy Tomlinson, Glory Marie, Mashaka Alimasi, Godfrey Stephano, Mlela Juma, Hussein Juma, Baruana Juma, Abdallah Said, Roda Dominick and Milka Hyamubi.

The use of human test subjects for this research was approved by the “University Research Ethics Committee” (UREC) of LJMU, with an approval reference of 17/NSP/008. Flights were performed according to the LJMU Operations Manual for UAVs as well as the Aeronautical Information Circular (AIC) number 5/17 (Pink 62) of 1 January 2017 and according to the Tanzania Civil Aviation Regulations (TCARs). 

\bibliography{report} 

\begin{thebibliography}{10}

\bibitem{Ceb15}
Ceballos, G., Ehrlich, P.~R., Barnosky, A.~D., Garc{\'\i}a, A., Pringle, R.~M.,
  and Palmer, T.~M., ``Accelerated modern human{\textendash}induced species
  losses: Entering the sixth mass extinction,'' {\em Science Advances}~{\bf
  1}(5) (2015).

\bibitem{wwf16}
WWF, ``Living planet report 2016. risk and resilience in a new era.,'' {\em WWF
  International, Gland, Switzerland}  (2016).

\bibitem{Schwarz08}
James~Schwarz, C., ``Advanced distance sampling: Estimating abundance of
  biological populations by buckland, s. t., anderson, d. r., burnham, k. p.,
  laake, j. l., borchers, c. l., and thomas, l.,'' ~{\bf 64},  997 (10 2008).

\bibitem{Fre12}
Fretwell, P.~T., LaRue, M.~A., Morin, P., Kooyman, G.~L., Wienecke, B.,
  Ratcliffe, N., Fox, A.~J., Fleming, A.~H., Porter, C., and Trathan, P.~N.,
  ``An emperor penguin population estimate: The first global, synoptic survey
  of a species from space,'' {\em PLOS ONE}~{\bf 7},  1--11 (04 2012).

\bibitem{McM14}
McMahon, C.~R., Howe, H., van~den Hoff, J., Alderman, R., Brolsma, H., and
  Hindell, M.~A., ``Satellites, the all-seeing eyes in the sky: Counting
  elephant seals from space,'' {\em PLOS ONE}~{\bf 9},  1--5 (03 2014).

\bibitem{Buc01}
Buckland, S., Anderson, D., Burnham, K., Laake, J., Borchers, D., and Thomas,
  L.,  [{\em Introduction to Distance Sampling: Estimating Abundance of
  Biological Populations}{\nolinebreak\hspace{0.1em}]}, Oxford University
  Press, United Kingdom (2001).

\bibitem{Franklin10}
Franklin, J.,  [{\em Mapping species distributions: spatial inference and
  prediction}{\nolinebreak\hspace{0.1em}]}, Cambridge University Press, United
  Kingdom (2010).

\bibitem{vermeulen13}
Vermeulen, C., Lejeune, P., Lisein, J., Sawadogo, P., and Bouche, P.,
  ``Unmanned aerial survey of elephants,'' {\em PLOS one}~{\bf 8}(2) (2013).

\bibitem{Wich18}
Wich, S. and Koh, L.,  [{\em Conservation Drones}{\nolinebreak\hspace{0.1em}]},
  Oxford University Press, United Kingdom (2018).

\bibitem{hodgeson18}
Hodgson, J.~C., Mott, R., Baylis, S.~M., Pham, T.~T., Wotherspoon, S.,
  Kilpatrick, A.~D., Segaran, R.~R., Reid, I., Terauds, A., and Koh, L.~P.,
  ``Drones count wildlife more accurately and precisely than humans,'' {\em
  Ecology and Evolution}~{\bf 9}(5) (2018).

\bibitem{Longmore17}
{Longmore}, S.~N., {Collins}, R.~P., {Pfeifer}, S., {Fox}, S.~E.,
  {Mulero-Pazmany}, M., {Bezombes}, F., {Goodwind}, A., {de Juan Ovelar}, M.,
  {Knapen}, J.~H., and {Wich}, S.~A., ``{Adapting astronomical source detection
  software to help detect animals in thermal images obtained by unmanned aerial
  systems},'' {\em International Journal of Remote Sensing}~{\bf 38},
  2623--2638 (Feb. 2017).

\bibitem{hambract18}
Hambract, L., Wich, S., et~al., ``The efficiency of thermal and rgb imaging for
  detecting humans with drones,''  in prep (2018).

\bibitem{rashman18}
Rashman, M., Steele, I., Burke, C., Longmore, S., and Wich, S., ``Adapting
  thermal-infrared technology and astronomical techniques for use in
  conservation biology,'' {\em SPIE proceedings} ,  in press (2018).

\bibitem{polar_bears}
Preciado, J., Rubinsky, B., Otten, D., Nelson, B., Martin, M., and Greif, R.,
  ``Radiative properties of polar bear hair,'' {\em Advances in Bioengineering,
  International Mechanical Engineering Congress and Exposition}  (2002).

\bibitem{burke18}
Burke, C., Rashman, M., Wich, S., Symons, A., Theron, C., and Longmore, S.,
  ``Optimizing observing strategies for monitoring warm-blooded animal species
  using uav-mounted thermal infrared cameras,'' {\em International Journal of
  Remote Sensing} ,  in press (2018).

\end{thebibliography}
\bibliographystyle{spiebib} 

\end{document}